\documentstyle{mn2}

\input{epsf}

\def\etal{{\rm et~al. }}
\def\spose#1{\hbox to 0pt{#1\hss}}
\def\simlt{\mathrel{\spose{\lower 3pt\hbox{$\mathchar"218$}}
     \raise 2.0pt\hbox{$\mathchar"13C$}}}
\def\simgt{\mathrel{\spose{\lower 3pt\hbox{$\mathchar"218$}}
     \raise 2.0pt\hbox{$\mathchar"13E$}}}
\newcommand{\calD}{{\cal D}}

\newcommand{\bea}{\begin{array}{c}}
\newcommand{\eea}{\end{array}}
\newcommand{\beq}{\begin{equation}}
\newcommand{\eeq}{\end{equation}}
\newcommand{\beqa}{\begin{eqnarray}}
\newcommand{\eeqa}{\end{eqnarray}}

\newcommand{\xibar}{\overline{\xi}}
\newcommand{\wbar}{\bar{w}}

\def\lim{{\rm lim}}

\def\aj{\,\rm{AJ~}}			
		
\def\apj{\,\rm{ApJ~}}			
		
\def\apjs{\,\rm{ApJS~}}

\begin{document}

\title[Testing deprojection algorithms]
{Testing deprojection algorithms on mock angular catalogues:
evidence for a break in the power spectrum}

\author[E. Gazta\~{n}aga and C.M. Baugh]
{ E. Gazta\~{n}aga $^{1,2}$ and {\rm C.M. Baugh}$^{3}$
\\
1. Institut d'Estudis Espacials de Catalunya, 	Research Unit (CSIC),
Edf. Nexus-104 - c/ Gran Capitan 2-4, 08034 Barcelona
\\
2. Department of Physics, Keble Road, Oxford OX1 3RH.
\\
3. Department of Physics, Science Laboratories, South Road, Durham DH1 3LE
}

\maketitle 
 
\def\mpc {h^{-1} {\rm Mpc}}
\def\mpcc {h^{-3} {\rm Mpc^3}}
\def\kpc {h^{-1} {\rm Kpc}}
\def\impc {h {\rm Mpc}^{-1}}
\def\and  {{\it {et al.} }}
\def\rmd {{\rm d}}

\begin{abstract}

We produce mock angular catalogues from simulations with different initial 
power spectra to test methods that recover measures of clustering in three
dimensions, such as the power spectrum, variance and higher order cumulants.
We find that the statistical properties derived from the angular
mock catalogues are in good agreement with the intrinsic clustering in the 
simulations. In particular, we concentrate on the detailed predictions for 
the shape of the power spectrum, $P(k)$. We find that there is good  evidence
for a break in the galaxy $P(k)$ at scales between $ 0.02 < k < 0.06 ~\impc$
using an inversion technique applied to the angular correlation function 
measured from the APM Galaxy Survey. For variants on the standard 
Cold Dark Matter model, a fit at the location of the break implies 
$\Omega h= 0.45 \pm 0.10$,  where $\Omega$ is the ratio of the total matter 
density to the critical density and Hubble's constant is parameterised as 
$H_{0}= 100~h{\rm km ~s}^{-1}{\rm Mpc}^{-1}$.
On slightly smaller, though still quasi-linear scales, there is a feature
in the APM power spectrum where the local slope changes appreciably,
with the best match to CDM models obtained for $\Omega h \simeq 0.2$.
Hence the location and narrowness of the break in the
APM power spectrum combined with the rapid change in its slope on
quasi-linear scales cannot be matched by any variant of CDM, including
models that have a non-zero cosmological constant or a
tilt to the slope of the primordial  $P(k)$. These results are independent
of the overall normalization of the CDM models or any simple bias that 
exists between the galaxy and mass distributions.

\end{abstract}


\section{Introduction}
Angular catalogues of galaxy positions provide us with powerful 
constraints on theories of structure formation in the universe.
The APM Galaxy Survey covers 4300 square degrees on the sky 
and contains over 2 million galaxies to a limiting apparent 
magnitude of $b_{J} \le 20.5$ (Maddox {\it et al}
1990a,b,c; 1996).
The shape of the angular correlation function measured 
from the survey at scales of $\theta > 1^{\circ}$  
indicates that the universe contains more structure on 
large scales than is predicted by the standard Cold Dark 
Matter scenario (Maddox {\it et al} 1990c).

Whilst this result is confirmed by the largest redshift 
surveys currently available ({\it e.g.} 
Efstathiou {\it et al.} 1990a, Saunders {\it et al} 
1991, Vogeley {\it et al } 1992, Fisher {\it et al} 1993, 
Tadros \& Efstathiou 1996), 
measurements of correlations in 3D catalogues are still 
noisy on scales $r \ge 10 \mpc$.
Only after the completion of the Sloan Digital Sky Survey 
(Gunn \& Weinberg 1995) will 
a 3D catalogue contain the same order of magnitude of objects 
as the APM Galaxy Survey. 

An additional complication in redshift catalogues is that the 
pattern of galaxy clustering is distorted by the peculiar 
motions of galaxies (Kaiser 1987).
This effect can boost the amplitude of the measured two-point 
correlations by anything between a factor of $1-2$ on large 
scales depending upon the survey and the method of analysis 
(see Table 1 in Cole, Fisher and Weinberg 1995). 

Whilst the next generation of redshift surveys will undoubtedly provide 
a wealth of new information that is not available in angular 
catalogues, it is important to take full advantage of the 
large number of galaxies and volume surveyed in the angular catalogues
(such as the APM and the parent catalogue for the Sloan Survey when 
it is complete) to extract information about the correlations on large scales.
Under certain assumptions, 
deprojection algorithms to recover the 3D correlations in real 
space have been developed for multi-point correlation functions
(e.g. Groth \& Peebles 1977, Fry \& Peebles 1978,  Peebles 1980), for
$J$-order cumulants 
of counts in cells (Gazta\~{n}aga 1994, 1995; hereafter G94 and G95) 
and for the power spectrum (Baugh \& Efstathiou 1993, BE93; 1994, BE94).
In this paper we present tests of these algorithms by 
constructing angular catalogues, with the same selection function 
and angular mask as the APM catalogue, from large 
numerical simulations.
We use sets of simulations that have been evolved to have a 3D 
power spectrum that matches closely the APM form recovered by BE93
and also simulations of CDM models. 

For our present purposes, we are concerned with testing for 
the presence of any systematic biases that arise from the 
projection process itself rather than from the actual 
construction of the APM Survey or corresponding angular catalogue
(some of these problems are addressed in detail by Maddox {\it et al} 1996)

The outline of the paper is as follows.
In Section 2 we describe the N-body simulations used to make mock 
catalogues in Section 3. We present and test the recovery methods in 
sections  4 and 5. 
In Section 6 we discuss our results and present the conclusions.

\section{N-body realisations of APM Galaxy Clustering}

\begin{figure}
{\epsfxsize=8.5truecm \epsfysize=8.5truecm 
\epsfbox[15 140 570 700]{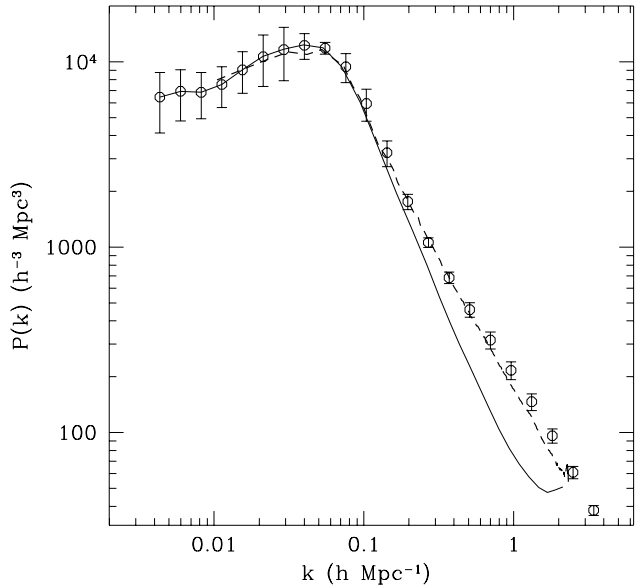}}
\caption[junk]{The power spectrum measured for APM Survey galaxies is 
shown by the open circles with the $1 \sigma$ scatter 
on the mean averaged over the survey split up into four zones.
The solid line shows the linear power spectrum estimated from this 
as described in the text.
The dashed line shows the power spectrum 
measured from the evolved simulation, APM2(a).}
\label{fig:pk}
\end{figure}

A discussion of the production of evolved N-body simulations 
that have the same power spectrum as that measured for 
APM Survey galaxies (BE93, BE94) is given in 
Baugh \& Gazta\~{n}aga (1996; BG96).
In this Section, we briefly summarize the approach taken and list the 
parameters of the N-body simulations that are used to make mock APM 
catalogues in Section \ref{s:mockapm}.

\begin{figure*}
\begin{picture}(300,650)
***Missing because of space restrictions in astro-ph.
Find it in: http://star-www.dur.ac.uk/~cmb/project.html
\end{picture}
\caption[junk]{
A comparison of the evolved density field in the 
APM2(a) (top) and CDM2(a) (bottom) simulations, 
which were started with the same random phases.
The density is binned on a $256^3$ grid and is smoothed with a Gaussian 
filter to blur the pixels.
The greyscale shows the logarithm of the density.
The slices are $3 \mpc $ thick and $600 \mpc$ square.}
\label{fig:slice}
\end{figure*}

\begin{table}
\begin{center}
\caption[dummy]{Simulation parameters. The third column gives the 
dimension of the FFT mesh used in the long range force calculation.
The last column gives the softening length of the gravitational 
force, $\epsilon$ 
in $h^{-1}$ {\it kiloparsec} units.}
\label{tab:param}
\begin{tabular}{lccccc}
\hline
\hline  
run         & number    & mesh    & $L_{box} $ &  $\epsilon$   \\ 
            & of particles         &         & $(\mpc) $  &   $(h^{-1}{\rm kpc}) $
 \\ \hline
APM1        & $160^{3}$ & $256^{3}$  & 440 &  115     \\
APM2(a)     & $200^{3}$ & $256^{3}$  & 600 &  156                 \\ 
APM3(a)-(e) & $126^{3}$ & $128^{3}$  & 400 & 104                \\
SCDM2(a)     & $200^{3}$ & $256^{3}$ & 600 & 156                 \\ 
SCDM3(a)     & $126^{3}$ & $128^{3}$ & 400 & 104                 \\ 
LCDM3(a)     & $126^{3}$ & $128^{3}$ & 400 & 104                 \\ 
\hline
\label{tab:prop}
\end{tabular}
\end{center}
\end{table}

The first step is to estimate the linear power spectrum from the 
measured power spectrum of APM Survey galaxies. 
This requires assumptions to be made about the cosmological 
model and the form of the bias, if any, between fluctuations 
in the light and the mass distributions (Kaiser 1984). 
In this paper we consider a spatially flat universe with the 
critical density $\Omega=1$ and zero cosmological constant.
We assume that there is no bias between light and mass, {\it i.e.} that 
light traces mass, for simplicity. 
The validity of this assumption is not important for the purposes of 
this paper, which are to generate a particular 
distribution of points in three dimensions and to 
determine how well the N-point correlations in 
3D can be recovered from a projected catalogue.
There is evidence that the relative bias between 
mass and light is small on large scales from the hierarchical scaling 
of higher order moments of galaxy counts in the APM Survey (G94), 
though this does not appear to be the case on smaller scales (BG96).

The linear power spectrum is obtained from the evolved power spectrum 
using the transformation of Jain, Mo \& White (1995). 
This transformation is based upon a suggestion by Hamilton etal. (1991) 
that a universal form exists relating the linear and nonlinear 
correlation functions.
The method was extended to power spectra by Peacock \& Dodds (1994) and 
modified by Jain \and (1995) to cope with steep power law 
fluctuation spectra, $P(k) \propto k^{n}$, with $n < -1$.
We have found that the formula of Jain etal. gives more 
self-consistent results for the $n \sim -2$ linear power spectra discussed 
here than the revised formula given by Peacock \& Dodds (1996).
  
The linear to nonlinear transformation is given by 
\begin{eqnarray}
\Delta^{2} (k_{NL})/b(n) &=& f_{NL} [ \Delta^{2}_{L} (k_{L})/b(n)] \\
k_{L} &=& [ 1 + \Delta^{2}_{NL} (k_{NL}) ]^{-1/3} k_{NL},
\end{eqnarray}
where the subscripts $L$ and $NL$ refer to linear and nonlinear 
respectively and $\Delta (k) = 4 \pi k^{3} P(k) / (2 \pi)^{3}$ is 
the fractional variance of the density field in bins of $ \ln k$.
The factor $b(n) = [(3 + n)/3]^{1.3}$ is a function of the 
effective spectral index of the density fluctuations, defined 
as the local slope of the linear power spectrum at the scale 
on which the variance is unity.
Using the functional form for the inverse of $f_{NL}$ given by equation 7(b) 
of Jain \and (1995), the linear power spectrum corresponding 
to the measured APM galaxy power spectrum can be calculated iteratively.
The linear APM power spectrum is shown by the solid 
line in Figure \ref{fig:pk}, with the measured APM galaxy 
power spectrum shown by the open circles. 
The errorbars show the $1 \sigma$ scatter in the mean from 
averaging over the APM Survey spilt up into four zones (BE93, BE94).
The linear APM power spectrum is smoother than the measured spectrum 
and is better fitted by a simple analytic form; for $k<0.6 \impc$
\begin{equation}
P_{APM}(k) \propto {k\over \left[1+(k/k_c)^2\right]^{3/2}},\label{pk}
\end{equation}
with $k_c\approx 150\ H_0/c$ (BG96).

The linear APM power spectrum is used to generate the initial 
density fluctuations in a N-body simulation. The simulation is 
evolved until the variance measured in spheres of radius $
30  \mpc$ matches that in the APM Survey. 
Several sets of simulations with APM initial conditions 
are used in this paper.
APM1 consists of one simulation with $160^3$ particles 
in a $440 \mpc$ box.
APM2 has one realization with  $200^3$ particles in a $600 \mpc$ box
and APM3 is an ensemble of five simulations (a)-(e), 
with half as many particles as APM1 and a slightly smaller box.
The parameters of the simulations are listed in Table \ref{tab:prop}.
The power spectrum of the evolved simulation APM2(a) is shown 
by the dashed line in Figure \ref{fig:pk}.
The evolved power spectra give a very close match to the 
measured APM power spectrum.
In all cases, we generate the initial conditions using a FFT on 
a $256^{3}$ potential grid ($N_g=256$).
The softening length of the APM3 and CDM3(a) runs was adjusted to be 
comparable to that used in the APM1 run.
All simulations were run using the ${\rm P}^{3}{\rm M}$ 
particle-particle/particle-mesh code of Efstathiou \and (1985).

We have run two simulations with the same random phases as APM2(a)
and APM3(a), but with the standard CDM power spectrum, 
$\Omega=1$ and $h=0.5$; these are called SCDM2(a) and SCDM3(a). 
We also ran another CDM simulation with the same random phases 
as APM3(a) and SCDM3(a), but with a low density parameter and 
a nonzero value of the cosmological constant, 
$\Omega=0.2$, $\Lambda=0.8$  and $h=1$,
which is called  LCDM3(a).
The initial density field in the CDM simulations is set up using the
transfer function of Bond \& Efstathiou (1984) for a universe with
baryon density  $\Omega_{B} = 0.03$.
This transfer function can be expressed in terms of a
parameter $\Gamma = \Omega h$ (Efstathiou, Bond \& White 1992);
note that this definition of the shape parameter $\Gamma$ is
relative to a model with $\Omega_{B}=0.03$ and differs
slighty from that adopted by Peacock \& Dodds (1994).
In all cases we have run the CDM simulations so that the
linear variance on scales of $8 \mpc$ is 
$\sigma_8 \simeq 0.84$ (note that this value does not take into
account any evolution in the clustering, and corresponds to
clustering at the mean redshift in the APM, e.g. G95).
The SCDM simulation has more power on small scales and 
less power on large scales than the APM run.
This can be seen in a comparison of the particle 
distributions from APM2(a) and 
SCDM2(a) shown in Figure \ref{fig:slice}.
The figure shows a slice from the simulation box, after the particle 
density has been tabulated on a $256^3$ grid and smoothed on small 
scales with a Gaussian filter.
The slice shown is $\sim 3 \mpc$ thick and $600 \mpc$ square.

\section{Mock APM maps}
\label{s:mockapm}

We transform the N-body simulation into a mock APM catalogue 
of angular positions by the following steps:

\begin{enumerate}
\item Select an arbitrary point in the simulated box to be the local 
`observer'.
\item Apply the APM Survey angular mask, including plate shapes and holes.
\item Include a simulated particle at coordinate distance $x$ from the observer
with probability given by the selection function $\psi(x)$.
\end{enumerate}

The discreteness of the density field in the N-body simulations 
means that the final maps have a slightly lower density than 
the real APM map. 
The total number of particles is about 
$8 \times 10^5$ compared with $1.3 \times 10^6$ galaxies
in the APM Survey to the same apparent magnitude limit. 
This introduces additional shot-noise in the measurements which
is corrected in the standard way (e.g. G94).
The simulations use a periodic box, so we replicate the box to cover
the total extent of the APM volume (over $1200 \mpc$,
beyond were the expected number of galaxies is of order unity).
By comparing the results from different box sizes we have 
verified that this replication of the box does not introduce 
any spurious correlations on large scales.

\subsection{The selection function}

\begin{figure}
\centering
\centerline
{\epsfxsize=7.truecm \epsfysize=7.truecm 
\epsfbox{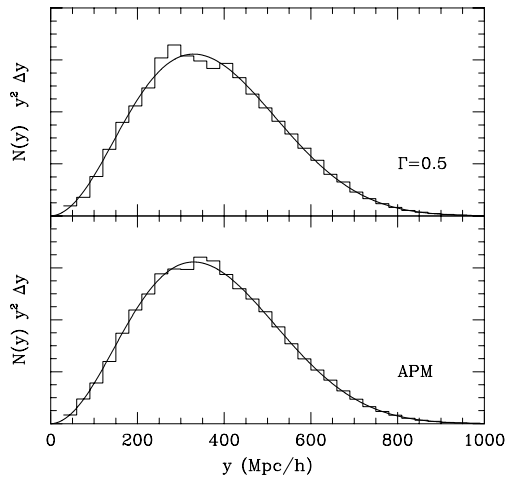}}
\caption[junk]
{Comparison of the theoretical (smooth curve) and measured 
counts (histogram) in radial shells for two mock catalogues 
made from the simulations SCDM3(a) (top) and APM3(a) (bottom).}

\label{nzdz}
\end{figure}

The selection function $\psi(x)$ is the normalized probability 
that a galaxy at coordinate distance $x$ is included in the catalogue. 
This probability is proportional to the estimated number of galaxies
at this coordinate:

\beq
\psi(x) = \psi^* \int_{q_1(x)}^{q_2(x)} ~dq~\phi(q) \label{psi}
\label{sel}
\eeq
where $\psi^*$ is adjusted so that the probability integrates 
to unity over the sample.
$\phi(q)$ is the luminosity function and $q_1(x)$ and $q_2(x)$ are the
scaled luminosities corresponding to the lower and upper limits
in the range of apparent magnitudes used 
to build the galaxy sample or catalog 
under study. In our case these are $b_J=17$ and $b_J=20$ respectively.
G95 constructed a $\chi^2$ test to find contours
of the values of the luminosity function parameters 
that best fit observational constraints on the luminosity 
and redshift distribution; 
the redshift evolution of the luminosity function 
as parameterised as $\phi^{*} = \phi_{0}^{*}( 1 + \phi_{1}^{*}z)$; 
$\alpha = \alpha_{0} + \alpha_{1}z$ and $M^{\*} = M_{0}^{\star} 
+ M_{1}^{\*}z$.
Here we use the best fit parameters obtained by G95: $\phi_{1}^{*} \sim 0$,
$\alpha_{1} = -4$ and $M^{1}_{*} = -2$ and the zeroth-order values 
of Loveday etal (1992): $\phi_{0}^{*} = 0.0112 h^{3} {\rm Mpc}^{-3}$,
$M_{0}^{*} = -19.73$, $\alpha_{0} = -1.11$.  
BE93 proposed a functional form for the
redshift distribution $N(z)$, discussed below in 
Section \ref{s:dep_pk}.
This $N(z)$ distribution gives very similar results 
for the selection function.

Figure \ref{nzdz} shows a comparison between the expected 
number of galaxies, $n(y) y^2 \Delta y$, at different radial
depths (in comoving coordinates $y$) given by the 
input selection function compared to the measured counts
for two different mock catalogues.

\begin{figure*}
\centering
\centerline
{\epsfysize=22.truecm 
\epsfbox{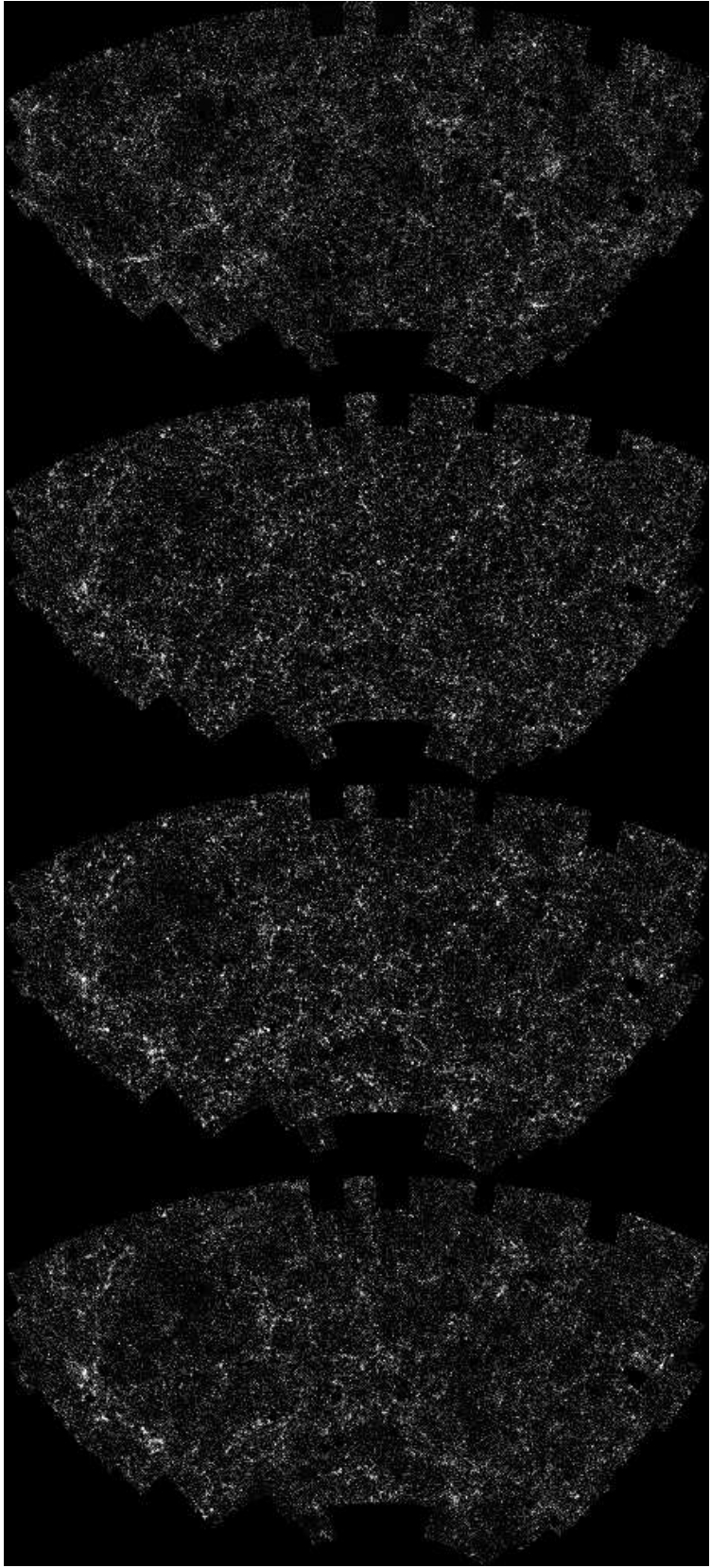}}
\caption[junk]
{Comparison of equal area projections of 
maps made from simulated catalogues with the real APM Galaxy 
Catalogue (Maddox \etal 1990a,b,c) (top).
The surface density of galaxies is represented by a greyscale, 
with the densest regions being the brightest.
In each map, the same total number of galaxies and the 
same greyscale calibration are used.
The maps extend about $120$ degrees in RA and 60 
degrees in DEC, covering about $20\%$ of the southern galactic cap, 
with a mean depth of $400 \mpc$.
The $185$ overlaping square
UK Schmidt plates in each map
correspond roughly to $5$ degrees on a side.
All maps have similar amplitudes of fluctuations ($\wbar_2$)  
at 1 degree. 
>From top to bottom we show the real APM Survey, the standard CDM map 
made from SCDM3(a), a lambda-CDM map made from LCDM3(a), and a 
mock APM map made from a simulation (APM3(a)) evolved to 
match the power spectrum of APM galaxies.}
\label{maps}
\end{figure*}

\subsection{Equal area projection maps}

We have made equal area projected maps from the mock catalogues.
To facilitate a comparison between the maps made 
from the different simulations and the map of the APM Galaxy 
Survey, the maps have been turned into greyscale plots 
shown in Figure \ref{maps}. 
Grey intensity increases as a low power ($\simeq 0.1$)
of the point density.
The mock catalogues are from the same realization of the 
random seeds, and therefore have the same fluctuations in the 
same places but with different amplitudes, given by the 
difference in the initial power spectrum and its 
subsequent non-linear evolution.
In the notation of Table \ref{tab:param} these maps 
are from the simulations APM3(a), SCDM3(a) and LCDM3(a).
The real APM map has been diluted to show the same mean
surface density. 
The angular correlations are given in Figures \ref{w2apm} and \ref{w3apm}.
A visual comparison shows that the SCDM
model does not have as strong large scale fluctuations 
as the APM map, which is confirmed by Figure \ref{w2apm} 
(as found earlier by Maddox \etal 1990c). 
The SCDM distribution is quite smooth on the largest scales. 
One can also see how both CDM models have 
larger fluctuations on the smallest scales in these maps, 
showing a distinctive granulation in grey scale. 
The mock APM map is the closest of the models to the 
real catalogue, as expected from the very good agreement 
in the variance ({\it cf} Figure \ref{w2apm}).
Of course, the locations of individual structures in the 
real and mock APM maps do not coincide.
Any statistical differences are due to differences in
higher order correlations.

\subsection{Angular correlations}

\begin{figure}
\centering
\centerline
{\epsfysize=8.5truecm 
\epsfbox{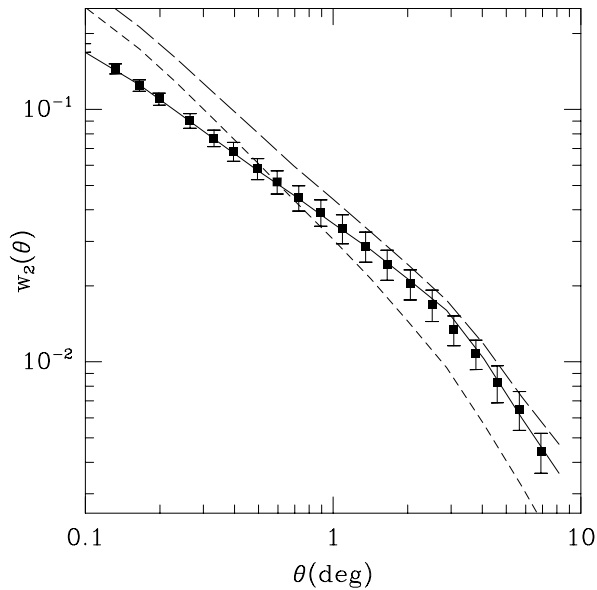}}
\caption[junk]
{Comparison of variance of angular counts-in-cells, $\wbar_2(\theta)$, 
in the simulated mock catalogues with the results in the real
APM data (symbols with errorbars).
A single realization of the maps made from the 
SCDM3(a), LCDM3(a) and APM3(a) simulations,  
normalised to a linear variance of $\sigma_8=0.84$, 
are shown as short-dashed, long-dashed and continuous lines. 
}
\label{w2apm}
\end{figure}

\begin{figure}
\centering
\centerline
{\epsfysize=8.5truecm 
\epsfbox{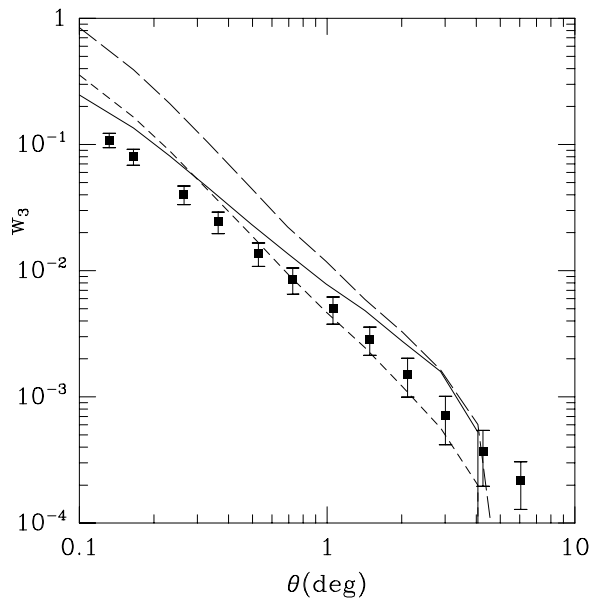}}
\caption[junk]
{Comparison of the skewness $\wbar_3$ 
of angular counts-in-cells,  
in the simulated mock catalogues  with the results in the 
APM Survey (symbols with errorbars). 
The SCDM3(a), LCDM3(a) and APM3(a) simulations,  
normalised to linear $\sigma_8=0.84$, 
are shown as short-dashed, long-dashed and continuous lines. 
}
\label{w3apm}
\end{figure}

Figure \ref{w2apm} compares the variance 
$\wbar_2 = <\delta^2>_c$, of angular fluctuations $\delta$ 
in cells of radius
$\theta$. 
The angular variance in the APM Survey is shown by the points 
with errorbars (G94).
The lines show the variance in the mock catalogues made from 
the SCDM3(a) (shot-dashed), LCDM3(a) (long-dashed) and APM3(a) 
(solid) simulations.
The CDM mock catalogues all have, by construction, 
the same linear theory normalisation $\sigma_8 \simeq 0.84$, 
but the non-linear $\sigma_8$ is, of course, slightly 
different in each case.
The simulated APM mock catalogue is slightly off the measured APM values
around $2-4$ degrees, this is also true for the mean of 
different realizations, and is probably due to slight inaccuracies 
in the match of the evolved power spectrum in the simulation to 
the measured galaxy power spectrum.
Note that all catalogues have similar power at around 1 deg.
As expected the LCDM model matches well the shape of the variance
at larger scales while the SCDM does not have enough large scale power.
At smaller scales the CDM models have too much power, as noted 
previously (e.g. Efstathiou \and 1990b, G95, BG96).

Figure \ref{w3apm}, shows  the skewness or third order reduced cumulant
$\wbar_3 = <\delta^3>_c$, for the same single realization of 
each model compared to the APM measurements. 
The errors show that at scales bigger than 1 degree
there are very large sampling fluctuations in $\wbar_3$. 
This is more dramatic in single realizations of each mock 
catalogues which cover a smaller volume than
the real APM. It is therefore dangerous to draw any conclusions, 
from this figure alone, at scales $\theta> 1$ deg. 
By comparing different realizations, we note that the mean 
at  $\theta> 1$ deg. comes closer to the APM observations.


In the following two Sections, 
we first give a brief review of the deprojection algorithms,
we then estimate the three dimensional statistics from the 
full simulation box i.e. the $\xibar_J$ using the counts in cells
method for the whole simulation box (as in Baugh, Gazta\~naga
\& Efstathiou 1995) or the power spectrum.
The two dimensional measurements of clustering, the 
$\wbar_J$ or the angular correlation function, are estimated 
from the mock catalogues (as in G94)
and the deprojection algorithms described 
in the previous Section are applied.

\begin{figure}
\centering
\centerline
{\epsfxsize=8.truecm 
\epsfbox{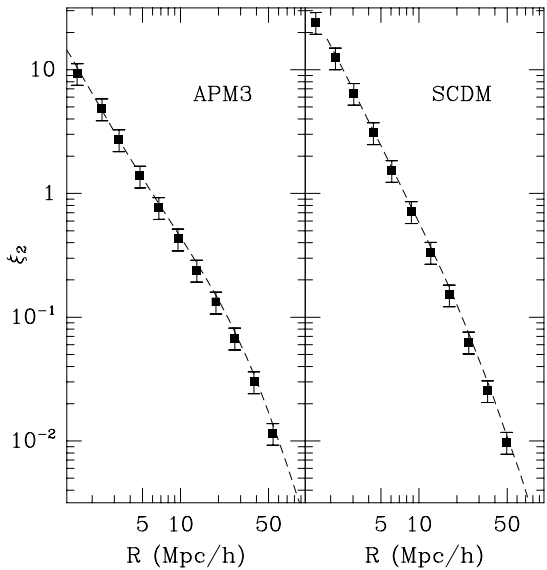}}
\caption[junk]
{Comparison of variance of counts-in-cells, $\xibar_2(R)$, 
in the simulated box (lines) with the results inverted from the
corresponding angular mock catalogue (points). The right panel
corresponds to the $\Gamma=0.5$ CDM model SCDM3(a), 
while the left panel shows
the results from the APM3(a) mock catalogue.}
\label{ix2}
\end{figure}

\section{Recovery of the moments of counts in cells.}

Here we use a simple method for recovering the 3-D variance,
$\xibar_2(R)$, and higher order reduced moments, $\xibar_J(R)$, from the
2-D correlations, $\wbar_J(\theta)$. This method was 
introduced and applied to the APM Galaxy survey in G94, G95, where a 
full description can be found.

In a scale-invariant model $\xibar_2 \propto R^{-\gamma}$
 with slope $\gamma$,
we can use the  expressions  in G95
to relate the estimated angular amplitudes
to the underlying three dimensional amplitudes,
i.e. $\sigma_8^2\equiv \xibar_2(R=8)$ 
and $S_J \equiv \xibar_J/\xibar_2^{J-1}$.
Here we consider a distribution that is not exactly scale-invariant
but has a slope $\gamma$ which is a slowly varying
function of scale. We call this a quasi-scale-invariant model
(see G95). It is then possible 
to apply a local inversion at each scale.
In principle the correlations on all scales $R$ contribute
to the correlations on angular scale $\theta$, but
because the sample has a finite depth, $\calD$,
there is a characteristic scale $R \simeq \calD \theta$.
In our analysis we relate 
angular scales $\theta$  to 3-D scales
using $R= \calD \theta$, where $\calD$ 
is the estimated distance which corresponds
to the mean redshift of the sample (see also Peebles 1980).
Although there is some ambiguity as to what the best definition of
$\calD$ should be, in the scale-invariant regime, we
find that the estimated amplitudes of $\xibar_J$ are insensitive to
changes in our chosen value of $\calD$ .

Thus at each given given scale $\theta$ with local slope $\gamma
=\gamma(\theta)$, we use the scale invariant expressions 
to relate the estimated local angular amplitudes
to the underlying three dimensional values.
This results in an estimation
for $\xibar_J$ as a function of the scale $R= \calD \theta$. 
This model was used in G94 and G95
to recover the 3D correlations in the APM Survey.

\subsection{Test of the variance}

Figure \ref{ix2} shows the inversion of $\xibar_2(R)$
from a standard $\Gamma=\Omega h = 0.5$
CDM mock angular catalogue (right panel), and for a  mock APM catalogue
(left panel) compared to the corresponding variance  $\xibar_2(R)$
estimated directly in the 3-dimensional simulation [e.g. SCDM3(a) and
APM3(a)].
The variance recovered from the angular 
distribution is a very good match to the
variance measured from the full simulation. 
There is a slight disagreement at scales around 
$R \sim 20 \mpc$, where there is a rapid change in the slope, as expected,
but the discrepancies are within $1\sigma$.

\begin{figure}
\centering
\centerline
{\epsfxsize=8.truecm 
\epsfbox{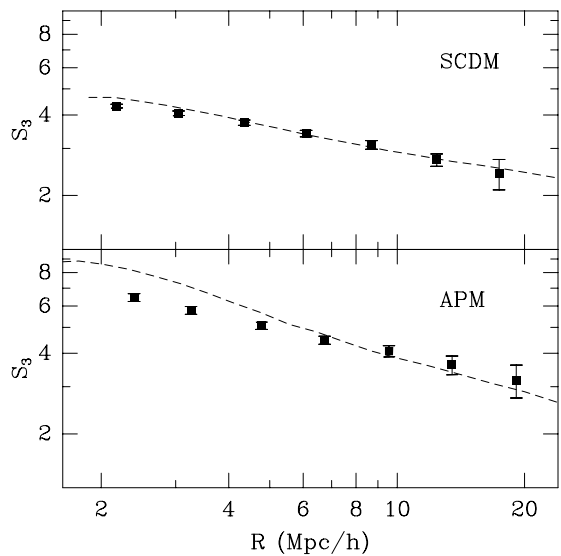}}
\caption[junk]
{Comparison of the skewness, $S_3(R)$, 
in the simulated box (lines) with the results inverted from the
corresponding angular mock catalogue (points). 
The top panel shows the
$\Gamma=0.5$ SCDM3(a) model, while the bottom panel corresponds
to the APM3(a) simulation.}
\label{is3}
\end{figure}

\subsection{Test of higher order moments}

The simulations we use have values of $S_J$ which show
a small variation with scale, e.g.  $S_3 \propto R^{-\alpha}$,
with $\alpha \simeq 0.1$. 
This indicates that strictly speaking neither the scale-invariant
nor the quasi-scale-invariant models should be used, as
$S_J$ should be constants in the hierarchical model.
Nevertheless, we still find  reasonable agreement from the 
inversion when we compare local values of $S_J$.

\begin{figure}
\centering
\centerline
{\epsfxsize=8.truecm
\epsfbox{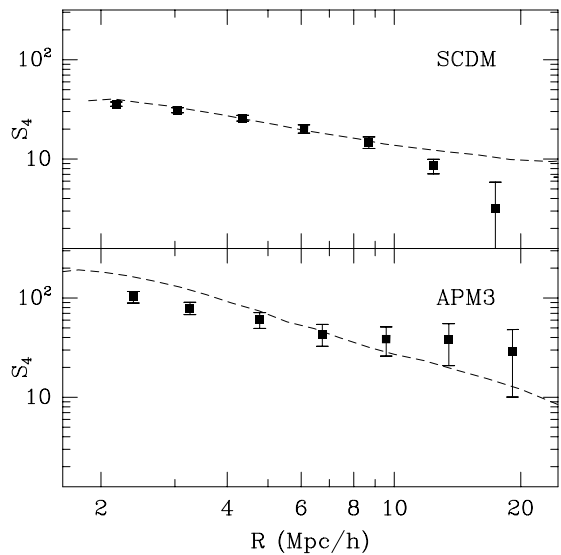}}
\caption[junk]
{Comparison of the kurtosis, $S_4(R)$, 
in the simulated boxes (lines) with the results obtained 
from the angular catalogues (as in Figure \ref{is3}).}
\label{is4}
\end{figure}

Figures \ref{is3} and \ref{is4} show the inversion of $S_3(R)$ and $S_4(R)$
from the standard $\Gamma=0.5$
CDM mock angular catalog and for the APM-like mock catalogue
compared to the corresponding amplitudes
estimated directly in the 3-dimensional simulated box. 
At scales $20 \mpc >R > 6 \mpc$, the amplitudes recovered from 
the angular distribution are in good agreement with 
the original amplitudes.  At larger scales, 
sampling fluctuations are very large,
whereas at smaller scales, there are some systematic
differences which seem more important for the APM model, which has a
steeper power spectrum.
As expected, the inversion method seems to work better 
for distributions where the 
$S_J$ are closer to being  constants, e.g. SCDM.
Note that the measured amplitudes $S_J$ in the APM are  
closer to a constant than either of the models we study here
(G94,G95) and one would then expect an even 
better agreement in this case.
The discrepancies at small scales could also be due in part
to shot-noise in
either the angular or the 3-dimensional distribution. 

As pointed out in Gazta\~{n}aga \& Bernardeau (1997),
there are several effects that make this type of comparison difficult.
First, volume and boundary effects are important on scales
$\simgt 2$ deg and tend to produce smaller values of the 
projected amplitudes $s_3$ 
and $s_4$. Second, the simple hierarchical
model for projections commonly used in the literature
(e.g. by Groth \& Peebles 1977, Fry \& Peebles
1978) is not accurate on 
quasi-linear scales, as indicated in Bernardeau (1995). 
These two effects compete with each other
and it is not clear how the projection model should be improved 
to allow a better reconstruction.

\section{Recovery of the power spectrum}
\label{s:dep_pk}

BE93, BE94 developed an iterative technique 
to numerically invert Limber's (1954) equation which relates a 
measure of clustering in 2D to an integral of the 3D power 
spectrum multiplied by the survey selection function.
BE93 used the measured angular correlation function of the 
APM Survey, $w(\theta)$ to obtain an estimate of the 3D power 
spectrum, whilst the 2D power spectrum, $P_{2}(k)$ was used in BE94.
An estimate of the real space correlation function has also been 
made in the same way (Baugh 1996).

This algorithm for the numerical inversion of Limber's equation 
does not rely upon the initial form chosen for the power spectrum 
and can reveal features that would be difficult to parameterize in 
a simple way.
The technique is numerically stable, unlike the use of Mellin 
transforms which involve differentiation of noisy 
quantities (Fall \& Tremaine 1977), and it has been shown to 
rapidly converge to stable solutions (BE93).

The integral equation relating $w(\theta)$ to the 3D power 
spectrum, $P(k)$, is given by (BE93, see Peacock 1991 for 
the non-relativistic form)

\begin{equation}
w(\omega) = \int_{0}^{\infty} P(k) k g(k \omega) {\rm d} k,
\label{eq:inteq1}
\end{equation}
where the angular variable is $\omega = 2 \sin( \theta/2)$ and the 
kernel function is an integral over the survey selection function

\begin{eqnarray}
g (k \omega) &= \frac{1}{2 \pi} \frac{1}{({\cal N} \Omega_{s})^{2}} 
\int_{0}^{\infty} \frac{F(x)}{(1+z)^{\alpha}} 
\left( \frac{ {\rm d} N}{{\rm d} z} \right)^{2} \\ \nonumber
&\frac{{\rm d} z}{{\rm d}x}
J_{0} (k \omega x) {\rm d} z, 
\label{eq:gker}
\end{eqnarray}
where $F(x)$ depends upon the cosmological model 
(see Peebles 1980, \S56) and 
$\Omega_{S}$ is the solid angle of the survey.
The time evolution of the power spectrum is parameterised as 
$P(k,z) = P(k)/(1 + z)^{\alpha}$, where $\alpha=0$ corresponds to 
the pattern of clustering being fixed in comoving coordinates, 
which is the case we use in this paper. This is a necessary 
oversimplification as we have an observed quantity that is a function of 
only one variable. Furthermore, the median redshift of the APM Galaxy survey 
is $ z_{m} \sim 0.12$ and the corrections for redshift evolution are small.

The redshift distribution of survey galaxies is parameterised as 
(BE93) :

\begin{equation}
\left( \frac{{\rm d} N}{{\rm d}z} \right) {\rm d} z 
= \frac{3 {\cal N}(m) \Omega_{s}}{2 z_{c}^{3}}
z^{2} \exp(-\left( \frac{z}{z_{c}}\right)^{3/2} ) {\rm d} z
\label{eq:nzdz}
\end{equation}
with the median redshift given by:
\begin{equation}
z_{m} = 1.412 z_c = 0.016 (b_{J} - 17)^{1.5} + 0.046
\end{equation}
for apparent magnitudes $b_{J} \ge 17$.
This form was chosen to provide a fit to the 
redshift distribution in the Stromlo/APM 
survey (Loveday {\it et al}. 1992) and to the 
fainter surveys of Broadhurst {\it et al.} (1988) and of 
Colless {\it et al.} (1990, 1993).
Redshifts have now been measured for galaxies in the 
magnitude range covered by the APM Survey, $17 \le b_{J} \le 20$, 
(Ellis {\it et al} 1996) and the redshift distribution is in 
good agreement with the form that we have 
adopted (Efstathiou private communication).

The $r^{th}$ iteration of the Lucy algorithm gives an estimate of the 
data, of 

\begin{equation}
w^{r} (\omega_{i}) = \sum_{j} P^{r}(k_{j}) 
g(k_{j} \omega_{i}) k^{2}_{j} \Delta \ln k
\end{equation}
which is compared with the `true' data, $w^{0}(\theta)$ in order to 
generate a new estimate of the power spectrum:
\begin{equation}
P^{r+1}(k_{j}) = P^{r}(k_{j}) \frac{\sum_{i} 
\frac{w^{0}(\omega_{i})}{w^{r}(\omega_{i})} 
g(k_{j} \omega_{i}) \Delta \ln \omega}
{\sum_{i} g(k_{j} \omega_{i}) \Delta \ln \omega}.
\end{equation}
The summations have typically 60 logarithmic bins for the 
data and 30 logarithmic bins for $P(k)$ in the range $ 3\times 10^{-3} \le 
k \le 30 \impc$.

\begin{figure}[t]
\centering
\centerline
{\epsfxsize=8.8truecm \epsfysize=8.8truecm 
\epsfbox{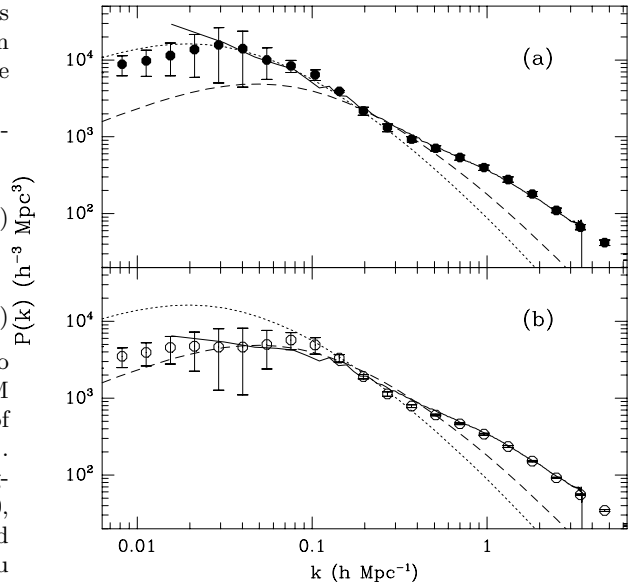}}
\caption[junk]{
Symbols with errorbars show the recovered 3D $P(k)$ from
the inversion of equation \ref{eq:inteq1} for maps made 
from the CDM simulations: a) LCDM3(a)  b) SCDM3(a). 
The dotted and dashed lines show for reference 
the linear power spectra of the LCDM and SCDM models respectively. 
The solid line in each case shows the nonlinear $P(k)$ measured from 
the simulation.
}
\label{fig:pkinv2}
\end{figure}

\subsection{Test of the Recovery of $P(k)$.}

In all cases, unless otherwise stated,
 we use the mean angular 2-point correlation function
and its variance in 4 individual disjoint zones
(shown in Figure 2 of BE94) to recover the power spectrum.
The results of the inversion of equation \ref{eq:inteq1} are 
illustrated in Figure \ref{fig:pkinv2}, for the two CDM 
models SCDM3(a) and LCDM3(a),  
which have power spectra with very different amplitudes 
and curvatures at a given wavenumber.
There is very good agreement in each case, up to the largest 
wavenumbers sampled in the simulation box in these runs, 
$ k \simeq 2\pi/L \simeq 0.015 \impc$.
In section \S\ref{sec:pkls} below we present a more detailed comparison
for larger scales.

\newpage

\section{Measuring the Break in $P(k)$.}

\subsection{Accuracy on large scales}
\label{sec:pkls}

\begin{figure}
\centering
\centerline
{\epsfxsize=8.truecm \epsfysize=8.truecm 
\epsfbox{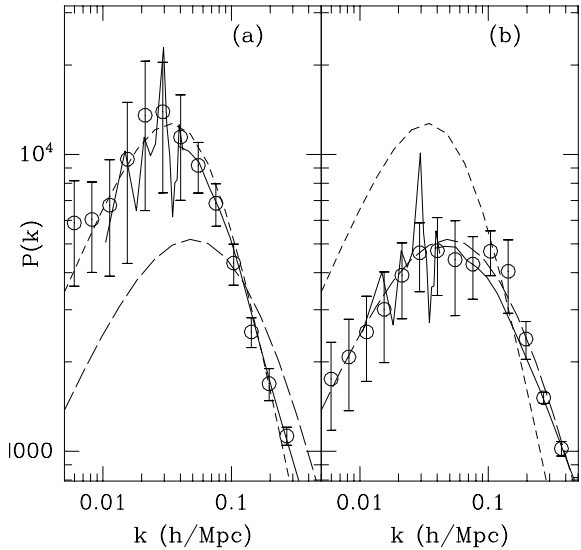}}
\caption[junk]
{Comparison of the recovered 3-D  $P(k)$ (points with errorbars), from
the angular catalogues, with the $P(k)$ from the corresponding 
3D simulation (continuous line).
The left hand panel shows the results for APM2(a) and the 
right hand panel shows SCDM2(a). 
The long and short dashed lines
correspond to the $\Gamma=0.5$ linear CDM $P(k)$
and the APM linear power spectrum respectively.}
\label{break}
\end{figure}

In order to show that the inversion method can accurately recover  
features at small wavenumbers (large scales), such as the break 
in $P(k)$, we now concentrate on the largest volume simulations, 
with a box size of $L=600 \mpc$ (see Table \ref{tab:param}). 
We study a single mock angular map from SCDM2(a) and APM2(a), 
with the same phase correlations and position for the observer.
Figure \ref{break} shows a comparison of the initial linear
$P(k)$ (dashed lines) with the non-linear $P(k)$ in 3D from the full  
box (continuous line) for both: (a) SCDM (right panel) and (b) the APM model 
(left panel).
Note how even the $P(k)$ measured in the 3D box has large fluctuations
at small $k$, and in particular a large spike at $k \simeq 0.03 \impc$.
This is due to the
small number of modes available to estimate $P(k)$ on these scales 
with the Fast Fourier Transform (FFT) technique. 
These estimates have not been 
averaged in bins and the initial spectrum amplitudes are drawn from a 
Gaussian distribution (and are not set equal to the mean). 
The mode where this spike is located only corresponds
to nx=2 ny=2 nz=0, so that there are few modes to average over.
This is seen in both the APM and CDM P(k) in this plot, 
due to these simulations being set up with the same phase distributions.

On large scales in Figure \ref{break} we plot the power spectrum
at the individual Fourier modes.
At large wavenumbers we have binned the 3D FFT estimation for clarity.
The recovered $P(k)$ from the angular two-point function 
(points with errorbars) shows excellent agreement with the 
original $P(k)$. 
Hence the volume of a single N-body box ($L=600 \mpc$) 
is large enough to simulate and recover large scales features 
in $P(k)$, even  at $k \sim 0.01 \impc$.

It is clear from this figure alone that there is a significant 
measurement of the break of the power spectrum. 
To make this more qualitative we now turn to the 
local slope of $P(k)$.

We want to focus in more detail on the shape of the power spectrum
by estimating the local logarithmic slope:

\beq
 n(k) \equiv {d \log{P(k)}\over{d \log{k}}}
\eeq

To do a numerical estimation we first bin the $P(k)$ data
and use standard polynomial interpolation and numerical differentiation 
(e.g. Press \etal 1992) in logarithmic space. The error in the slope
is obtained assuming no spread in $k$:

\beq
\Delta n(k) \simeq {{d \Delta P(k)/P(k)}\over{d \log{k}}}.
\eeq
This approach seems to work well in the mock maps and 
avoids spreading the 
systematic errors coming from the sampling variance,
which typically introduces a larger uncertainty in the amplitude
of the correlations than in their shape (see Figure 4 in
Baugh etal.  1995).

\begin{figure}
\centering
\centerline
{\epsfxsize=8.truecm \epsfysize=8.truecm 
\epsfbox{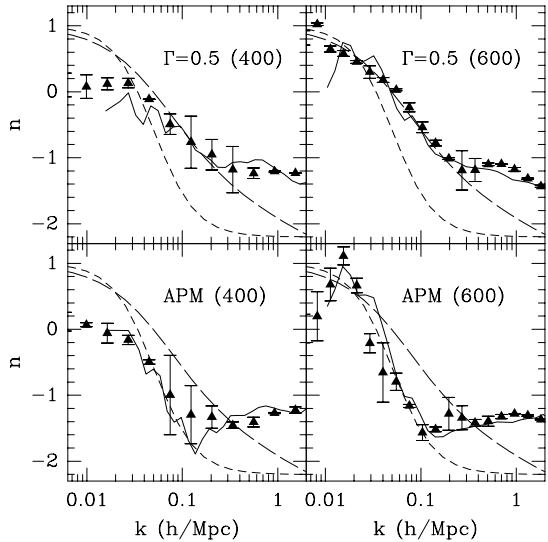}}
\caption[junk]
{Comparison of the recovered 3-D local slope $n(k)$ 
(points with errorbars), from the angular catalogues, 
with the slope in the corresponding 
3D power spectrum (continuous line). 
The long and short dashed lines
correspond to the $\Gamma=0.5$ linear CDM model
and the APM linear model. The left and right panels show the results in
the small ($400\mpc$) and large ($600\mpc$ boxes). The top and bottom 
panels correspond to SCDM and APM simulations respectively.}
\label{slope}
\end{figure}

Figure \ref{slope} shows the results for single realizations of
the SCDM and APM models for two different box sizes; 
SCDM2(a) and AMP2(a) with box size $L=600 \mpc$ and SCDM3(a) and 
APM3(a) which have a box size of $L=400 \mpc$.
The largest scales sampled in each pair of simulations correspond
 to wavenumbers 
of $ k \simeq 2\pi/L \simeq 0.01 \impc$ and $ k \simeq 0.015 \impc$ 
respectively for the  $L=600 \mpc$ and $L=400 \mpc$ boxes.
The smallest scales sampled are limited by the Nyquist frequency
of the FFT grid (of size $N_g$): $ k \simeq N_g\pi/L$, or for large enough
$N_g$ by the numerical resolution ($\epsilon$ in Table \ref{tab:prop}).

Figure \ref{slope} shows that the recovered slope matches closely 
that obtained directly in three dimensions, 
both in the non-linear ($k>0.2 \impc$)
and linear ($k<0.1 \impc$) regimes.
The particular realisations of the smaller boxes shown in the 
Figure \ref{slope} have a flatter slope on large scales 
in three dimensions than the
corresponding linear spectrum due to finite volume 
effects. 
This effect is also reproduced in the recovered slopes.

The break in $P(k)$ corresponds to $n=0$, and is well traced within the
errors in the larger boxes.

\subsection{Implications for the APM Power Spectrum.}

\begin{figure}
\centering
\centerline
{\epsfxsize=8.truecm \epsfysize=8.truecm 
\epsfbox{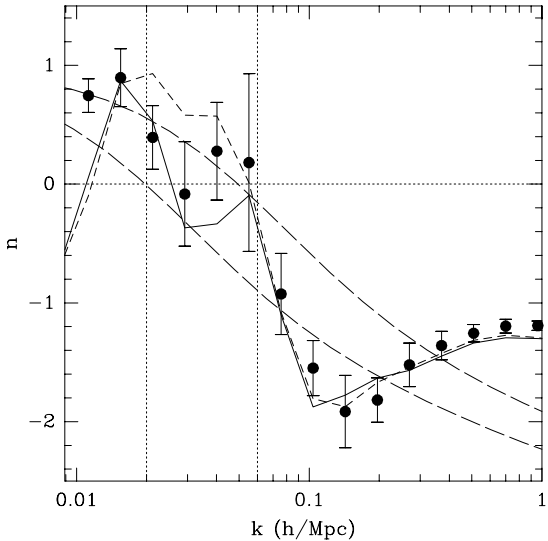}}
\caption[junk]
{The local slope of the power spectrum estimated from the APM catalogue. 
Symbols with errorbars correspond to the slope 
estimated in 4 individual disjoint zones (subsamples).
The continuous line is
from an inversion of the angular correlation 
function measured from the full APM map. 
The short dashed line corresponds to the inversion after
subtracting $10^{-3}$ from the angular correlation
function in the full APM map. 
The two long-dashed lines correspond to
linear CDM models with $\Omega h=0.5$ (top) and $\Omega h=0.2$
(botom). }
\label{apmsl}
\end{figure}

\begin{figure}
\centering
\centerline
{\epsfxsize=8.truecm \epsfysize=8.truecm 
\epsfbox{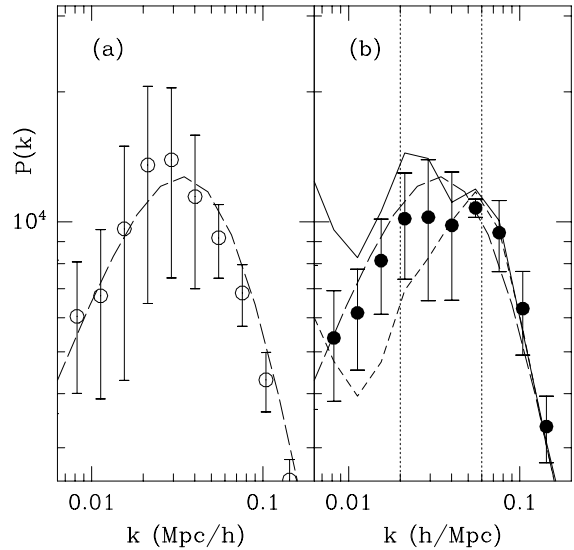}}
\caption[junk]
{Comparison of the recovered  $P(k)$ (points with errorbars), from
the angular catalogues with the linear APM model (long-dashed line). 
Panel (a) corresponds to a mock APM catalogue [APM2(a)]. 
Panel (b) shows the estimated APM $P(k)$ from 
measurements in the real galaxy catalogue.
In both cases the points and errorbars correspond to the mean and
variance in 4 individual disjoint zones (subsamples). The
continuous line in panel (b) corresponds to the inversion result 
obtained using the 
angular correlation function measured from  the full APM map.
The short-dashed line corresponds to the inversion result 
after subtracting an offset of $10^{-3}$ from $w(\theta)$.}
\label{breakapm}
\end{figure}

Table \ref{tab:apmpk} shows the values of $P(k)$
recovered  from the two-point correlations in the APM angular 
Galaxy Catalogue. These are essentially the same as in
Figure 7 of BE93, although there are small differences
corresponding to a different number of iterations in the
Lucy algorithm (chosen here to provide the minimum $\chi^2$
match to the angular correlation function).
Figures \ref{apmsl} and \ref{breakapm}
illustrate the  implications of our findings 
for the power spectrum recovered from the APM.
Figure \ref{apmsl} shows the reconstructed slope in the
APM Galaxy power spectrum, while Figure \ref{breakapm} shows 
the corresponding $P(k)$. Symbols with errorbars correspond to the 
mean and variance in 4 individual disjoint zones 
(shown in Table~\ref{tab:apmpk}). 
The break at $n=0$ is found to lie
between $k=0.02-0.06 \impc$ (between the vertical 
dotted lines in the Figure).
This can also be shown directly in Figure \ref{breakapm}, where
$P(k)$ shows a significant break on similar scales
(also bounded by dotted lines). Note that the errorbars
are comparable in the mock and the real catalogues.

The power spectrum recovered from the mock catalogues 
agrees well with the $P(k)$ measured from the unprojected simulation 
box, indicating that the volume of a single box ($L=600 \mpc$) is 
large enough to realize and recover a break on scales around 
$k \simeq 0.05 \impc$, without any finite volume effects.
Thus the volume traced by the APM Survey 
(which extends radially well beyond $600 \mpc$) 
is large enough to allow a measurement of the break in the 
power spectrum, $n=0$.

In an extensive analysis of the systematic errors involved in 
plate matching, Maddox {\it et al} (1996) have placed an upper limit 
of $\delta w(\theta) \sim 1 \times 10^{-3}$ on the likely 
contribution of the systematic errors to the angular 
correlations. 
In Figures \ref{apmsl} and \ref{breakapm}
the inversion result using the angular correlation function measured from 
the full survey is shown as a continuous
line. The short dashed lines in these figures show how 
this result for the power 
spectrum  changes when an offset of $10^{-3}$ is 
subtracted from the angular correlation function 
in the full APM map.
In principle, results from individual zones (symbols with errors in the
Figures) could  be affected more by the zone boundary than results from
the full survey, though on the other hand large scale noise 
from plate matching could be more important 
for the whole survey than for 
individual zones.  For the mock catalogues, 
the smaller size of individual zones 
does not seem to introduce important errors at the
scales under consideration  (e.g. Figure \ref{slope}). Thus, while
 there is no clear reason to prefer the estimate of the power 
spectrum made from the full survey to  
that made from the zones, the later is less likely to be affected by any large
scale plate matching errors.
As the variance from the different zones 
includes all the above sources of potential error, 
we take this estimation and variance as our best mean and
errors.

\begin{figure*}
\centering
\centerline
{\epsfysize=12.8truecm 
\epsfbox{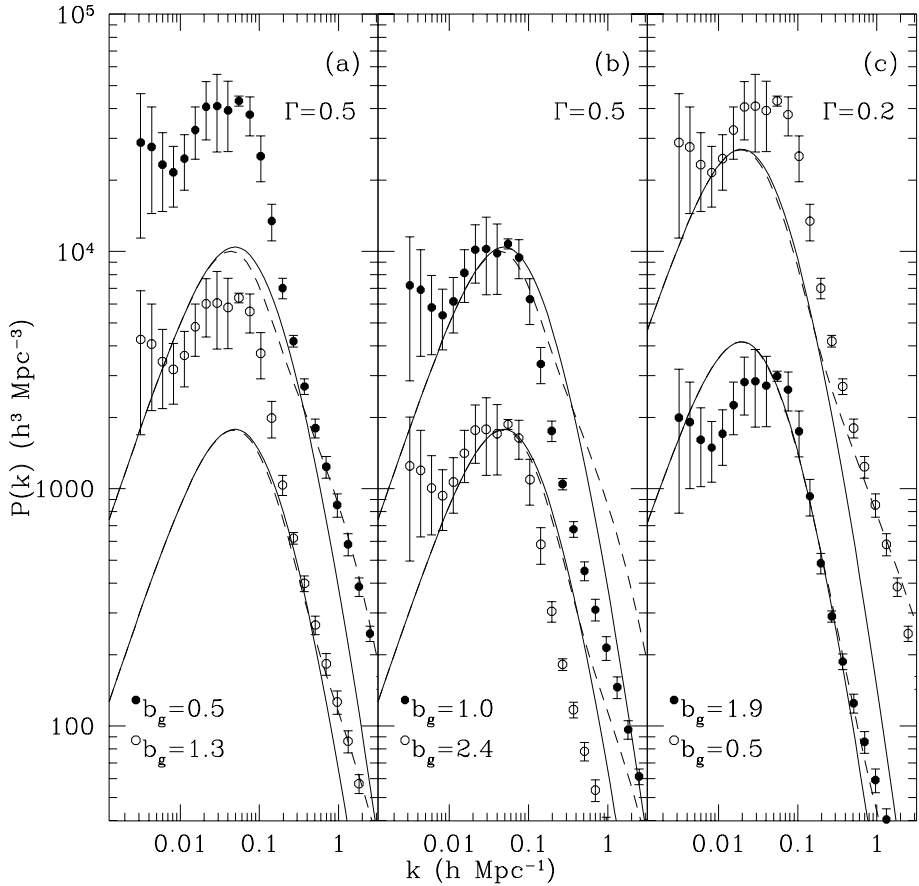}}
\caption[junk]
{The effects of nonlinear evolution on the shape of 
CDM power spectrum.
The solid lines show the linear theory power spectrum:
the lower and upper curves in (a) and (b) are
for normalisations of $\sigma_{8}=0.50$ and 
$\sigma_{8}=1.21$ respectively in the $\Gamma=0.5$ CDM model.
The lower solid curve in panel (c) shows a 
$\Gamma=0.2$ CDM model for a universe with
$\Omega=1$ and a Hubble constant of
$H_{0}=50 {\rm kms}^{-1}{\rm Mpc}^{-1}$ normalized to match the
COBE result with $\sigma_8=0.42$.
The upper solid curve in panel (c) corresponds to a open universe
with $\Omega=0.2$ and $H_{0} = 100 {\rm kms}^{-1}{\rm Mpc}^{-1}$,
with a normalisation that reproduces the abundance of rich clusters,
$\sigma_{8}=1.07$.
The dashed curves show the corresponding nonlinear power spectra;
in this case we have used the
transformation of Peacock and Dodds (1996),
which is expressed in a form that makes it 
readily appliciable to different cosmological models.

The APM galaxy power spectrum is shown by points with error bars, 
where we have divided by a bias parameter, $b_g$, squared 
(values indicated in the figure) to match 
the amplitude of the mass spectrum at different scales:
(a) the bias parameters have been chosen to match the 
amplitude of the mass $\Gamma=0.5$ CDM $P(k)$
on small scales $k \simgt 0.5$;
(b) the match is made to the amplitude of the mass 
$\Gamma=0.5$ CDM spectrum 
at the scale of the break in the APM power spectrum,
$k \simeq 0.05$; (c)  match to the amplitude of the mass 
$\Gamma=0.2$ CDM spectrum on small scales, $k \simgt 0.5$..
At small wavenumbers, $k < 0.01 \impc$, the estimate of the 
APM power spectrum is dominated by systematic and random errors in the 
catalogue}
\label{fig:pknonlin}
\end{figure*}

Figure \ref{fig:pknonlin} shows the effects of nonlinear 
evolution in the mass power spectrum for the $\Gamma=0.5$
standard CDM model and for two variants of SCDM with $\Gamma=0.2$.
Again, we use the form of the CDM power spectrum given by
Bond \& Efstathiou (1984), which is valid for a universe with a
small baryon density, $\Omega_{B}=0.03$, and we follow the
definition of $\Gamma= \Omega h = 0.5$ for SCDM adopted by
Efstathiou etal (1992).
For $\Gamma=0.5$ we show linear theory power spectra (solid lines) 
for two different normalisations to the variance in spheres of 
radius $8 \mpc$; the amplitude of temperature fluctuations in 
the microwave background gives a value $\sigma_8 \simeq 1.2$ 
(e.g. Stompor etal 1995, Bunn, Liddle  \& White 1996), 
whilst normalisation to reproduce the abundance of rich clusters requires  
$\sigma_8 \simeq 0.50 $,  virtually independent of the shape of the 
power spectrum for $\Omega=1$ (Eke, Cole \& Frenk 1996; White, Efstathiou, 
\& Frenk 1993).
The dashed lines give the corresponding predictions for the 
nonlinear spectra, using the transformation of Peacock \& Dodds (1996)
rather than Jain \and (1995), which is not so accurate for CDM models 
(see BG96).
The lower set of curves in Fig \ref{fig:pknonlin}(c) show a critical
density model with a Hubble constant
$H_{0}=50 {\rm kms}^{-1}{\rm Mpc}^{-1}$, but with the SCDM transfer function
altered by using $\Gamma = 0.2$. The normalisation of this curve matches the
COBE detection, with $\sigma_{8}=0.42$.
The upper curves in (c) are for an open model with density parameter
$\Omega=0.2$ and $H_{0}=100 {\rm kms}^{-1}{\rm Mpc}^{-1}$. In this
case, the model is normalised to reproduce the abundance of rich clusters
with $\sigma_{8} = 1.07$ (Eke etal 1996).

The APM galaxy power spectrum has been plotted 
to match the amplitude of the mass power spectrum, dividing the observed 
amplitude by a bias parameter squared.
In the (a) panel, the bias parameters have been chosen to match the 
galaxy power spectrum to the amplitude of the mass power spectrum at 
wavenumbers $0.3 \le k \le 10 \impc$, whilst in the (b) 
panel, the match is made at the scale of the break in the APM power spectrum.
Panel (c) shows the case for $\Gamma=0.2$ normalized to small scales.
This Figure demonstrates the basic problem of the CDM model; the shape of the 
power spectrum cannot be made to match to observed  galaxy power spectrum 
at both large and small scales, unless some complicated biasing 
prescription is invoked, in which the bias would need to vary significantly 
with scale.

\subsubsection{The effect of biasing}
\label{sec:bias}

The fluctuations traced by the galaxy distribution might be different,
or biased, from the underlying mass fluctuations (e.g. Bardeen et al 1986). 
We will argue here that the effect of this biasing is not important 
for the shape of the APM power spectrum at large scales.

Assume that the (smoothed) 
galaxy fluctuations $\delta_g$ are related to the mass $\delta_m$ 
fluctuations by a local transformation:  $\delta_g(x)= F[\delta_m(x)]$, 
and that this relation can be given as a Taylor
series: $F = b_1 \delta_m + b_2 \delta_m^2+ ...$. Then the
two-point function $\xi_2^g(r) \equiv <\delta_g(x)\delta_g(x+r)>$
on a scale $r$  will be just given by: 

\begin{eqnarray}
\xi_2^g(r) &=& b_1^2 \xi_2^m(r) + b_1 b_2 <\delta_m(x)\delta_m(x+r)^2> +
\nonumber \\
&~& + b_1 b_2 <\delta_m^2(x)\delta_m(x+r)> + \cdots
\end{eqnarray}

were all further terms are of order 4 or greater in $\delta_m$,
and therefore correspond to either higher order correlations,
$\xi_J$ with $J>2$, or higher powers in $\xi_2$. 
If $\delta_m$ is Gaussian or hierarchical
(as in the case for gravitational evolution) the
higher order correlations $\xi_J$ are at most of order
$\xi_2^{J-1}$. This means that at large scales, where $\xi_2<1$,
the first term is the dominant one, so that only the amplitude but not 
the shape of the two-point
statistics is changed by biasing. This effect
have been found in N-body simulations and toy biasing models 
(Weinberg 1994, Mo, Jing \& White 1997, Gazta\~naga \& Lacey 1997).

For the APM power spectrum a wavenumber around  
$k \simeq 0.1 \impc$, corresponds to
a top-hat radius of $R \sim \pi/k \simeq 30 \mpc$. 
For any reasonable biasing model relating galaxy fluctuations,
$\delta_g$, to the underlying  matter  fluctuations, $\delta_m$, the 
matter density fluctuations are very small around $R \simeq 30 \mpc$.
The independent constraints on the normalisation of mass fluctuations 
discussed above give values of around unity for the variance in spheres of 
radius $8 \mpc$.
To have rms fluctuations of order unity at $R \simeq 30 \mpc$ 
would imply $\sigma_8 > 3$.

Thus from the above arguments, the small variance on large scales,
$R \simgt 8 \mpc$, means that
it is reasonable to assume that the galaxy shape of $P(k)$ 
for $k < 0.1 \impc$ corresponds to the shape of the underlying linear matter
power spectrum. This argument, just based on the smallness of the 
variance and the hierarchical structure, can also be applied to
gravity, as the leading contribution to the correlation functions
in perturbation theory is indeed exactly given by a local transformation
(see Fosalba \&  Gazta\~naga 1997). This is clearly 
illustrated in  Figure \ref{slope}. By comparing the linear and
non-linear shape of $P(k)$, one case see that it
has not been changed significantly 
by gravitational evolution on scales where the rms fluctuations are small,
i.e. $k < 0.1 \impc$.

\begin{table}
\begin{center}
\caption[dummy]{Values of the estimated power spectrum $P(k)$ recovered from 
measurements in the (real) APM angular galaxy catalogue, corresponding
to the mean and error from the variance in 4 individual disjoint regions
in the catalogue.}
\label{tab:param}
\begin{tabular}{crr}
\hline
\hline  
$k$ & $P(k)$ & $\Delta P(k)$ \\
$\impc$ & $(h^{-3} Mpc^{3})$ & $(h^{-3} Mpc^{3}) $ \\
\hline
\hline
    0.0032  &      7198   &     4345 \\
    0.0043  &      6891   &     3278\\
    0.0060  &      5805   &     2126\\
    0.0082  &      5386   &     1543\\
    0.0113  &      6158   &     1620\\
    0.0155  &      8134   &     2026\\
    0.0213  &     10174   &     2803\\
    0.0292  &     10251  &      3682\\
    0.0401  &    9821  &      3232\\
    0.0551  &   10776  &       523\\
    0.0757   &     9440  &      1770\\
    0.104   &     6299  &       1383\\
     0.143   &     3358  &       590\\
      0.196   &     1754   &    173\\
      0.270   &     1048  &      58\\
      0.371   &      675  &     51\\
      0.509   &     451   &    41\\
      0.700   &     309   &     33\\
      0.961   &     214 &       24\\
       1.32   &    146  &     16\\
       1.81   &    96.7  &     8.8\\
       2.49   &    61.4  &     4.6\\
       3.42   &    38.1 &       2.4\\
       4.70   &    23.8  &      1.3\\
       6.46   &     15.1  &    0.7\\
       8.88   &     9.63  &    0.43\\
        12.2   &    6.19  &    0.25\\
       16.8   &    4.15  &     0.15\\
       23.0   &    3.05  &   0.10\\
       31.6   &    2.42  &   0.07\\
\hline
\label{tab:apmpk}
\end{tabular}
\end{center}
\end{table}

\subsubsection{Variations of CDM models}

\label{sec:varCDM}

A simple variation of CDM models is to introduce a tilt
in the initial power spectrum so that:
$P(k) = k^{n_0} T(k)$, where 
$T(k)$ is the transfer function (e.g. Bond \& Efstathiou 1984,
Bardeen \etal 1986) and $n_0$, is the primordial spectral index, 
$n_0 \neq 1$.
Unless the tranfer function $T(k)$
somehow depends strongly on $n_0$, the local slope of 
a given tilted CDM model is similar to that of the corresponding
standard scale invariant model (where $n_0=1$), given by 
$n = n_0 + dlog(T)/dlog(k)$, with the shift due to the tilted value of 
$n_0$. 
Thus, tilted models
can only scale up or down the CDM predictions
in Figure \ref{apmsl}, and therefore can not account for the APM 
observations.

 The measurement of the abundance of deuterium in high redshift 
 hydrogen clouds is provoking much debate in the 
 literature (e.g. Rugers \& Hogan 1996, Tytler, Fan \& Burles 1996). 
 Consequently the baryon density of the universe is uncertain and  
 possible values fall in a wider range than was previously accepted.
 In the limit of a high baryon density (i.e. $\Omega_{B} \sim 0.1$), 
 the power spectrum of the mass is modified. A full calculation of the 
 transfer function (e.g. Slejak \& Zaldarriaga 1996) indicates that the 
 high baryon density introduces features or `wiggles' into the shape of 
 the power spectrum on large scales (see also Goldberg \& Hamilton 1997 
 for a discussion of how these peaks could be used to constrain the value 
 of $\Omega_{B}$).
 We have used the CMBfast code of Slejak \& Zaldarriaga to compute the 
 shape of the power spectrum in a CDM universe with $\Omega_{B}=0.1$ and 
 $\Omega = 1$. 
 The resulting modification of the power spectrum compared with the 
 Bond \& Efstathiou (1984) transfer function for $\Omega_{B}=0.03$ is 
 insufficient to improve the agreement with the APM power spectrum.

\section{Conclusions}

The algorithms tested here successfully recover the power spectrum and 
higher order cumulants in three dimensions.
There are no systematic shifts or biases in the inferred correlations 
resulting either from the deprojection techniques or from the process of 
projecting the original particle distribution.
The 3D variance recovered from angular catalogues is in good agreement
with the input model, confirming the results in G95. 
For higher order correlations the deprojection method studied here, and
also used in G94, Gazta\~naga \& Frieman (1994), G95 and BG96,
seems to be adequate, 
at least for intermediate scales, $20 \mpc >R > 6 \mpc$, 
although one would in principle expect deviations
from the smiple hierarchical,
 according to perturbation theory
(Bernardeau 1995). A more detailed analysis of this point
is presented elsewhere (see Gazta\~{n}aga \& 
Bernardeau 1997).

It is possible to recover the detailed shape of the power spectrum  
with errorbars similar to those quoted by BE93. As pointed out
there (and also in G95), the uncertainties in the selection function do not
have much effect on the recovered shape.
The deprojection algorithm is able to distinguish sharp features, 
such as the one between $k \simeq 0.07-0.2 \impc$
shown in Figure \ref{apmsl}, first remarked upon by BE93.
For this range of wavenumbers, the best fitting CDM model 
has $\Omega h \simeq 0.2$, as pointed out by Efstathiou etal (1990b) and 
Peacock \& Dodds (1994).
However, the break in the power spectrum in this particular CDM model 
is broader and at a larger scale than the break in the APM power spectrum.

We have shown that the volume traced by the APM Survey 
is large enough to allow a significant measurement of the break in the 
power spectrum, $n=0$, as found on scales around 
$k \simeq 0.05 \impc$.
We have also shown (See Figure 14) that possible systematic errors 
involved in the APM plate matching lie within our estimated errors.

Peacock \& Dodds (1994) report a break in the power spectrum at 
a wavenumber of $k \simeq 0.03 \impc$ using spectra 
measured from a range of different surveys. 
The volumes mapped out by these surveys span a considerable range.
We have found that only our largest simulation boxes allow the break 
to be measured accurately, both in the direct estimation of the 
power spectrum in three dimensions and in the recovered spectrum obtained 
from the projected catalogue.
The size of the largest box we use, $L=600 \mpc$, is much greater  
than the median depth of any of the redshift surveys available to  
Peacock \& Dodds, indicating that finite volume effects could 
have altered the shape of the power spectra estimated from individual surveys 
on large scales 
(as found in Figure \ref{slope} for the $400 \mpc$ boxes).
Other sources of uncertainty in this type of compilation
include the different selection biases 
applied, the differences in the intrinsic luminosities of the 
objects selected in the catalogues
and the large sampling variance from the smaller surveys.
Furthermore, the linearisation process applied to the measured power 
spectra involves a correction for the distortion of the pattern of clustering 
by galaxy peculiar velocities (Kaiser 1987), which is both model and
catalogue dependent (e.g. Smith etal 1997).

The location of the break that we find in the 
galaxy power spectrum matches that found in power spectrum 
of galaxies clusters, both from a compilation based on the Abell 
catalogue (Einasto etal 1997) and from a cafefully selected redshift 
sample drawn from the APM Cluster catalgoue (Tadros 1996, Tadros \& Dalton 
1997, Dalton etal 1992).

The physical interpretation of the break at 
\beq
k_B \simeq ~0.05~ \impc \simeq ~150~ {H_0\over{c}},
\eeq
found in the APM is unclear. We have argued in section \S\ref{sec:bias}
 that 
the galaxy shape of $P(k)$ 
for $k < 0.1 \impc$ corresponds to the shape in the underlying linear matter
power spectrum. 
 For inflationary models with Cold Dark Matter (CDM)
 the break in the power spectrum at 
 wavenumber $k_B$ corresponds to the Hubble radius when the universe
 becomes matter dominated. 
 This is because the amplitude of fluctuations is frozen as they
 enter the Hubble radius during the radiation dominated era 
 (see Bond \& Efstathiou 1984, Bardeen \etal 1986).
 The wavelength of the Hubble radius at this epoch 
 is $\lambda_B \sim 10 (\Omega h)^{-1}
 \mpc$ (e.g. Kolb \& Turner 1990),
 where $\Omega$ is the total 
 matter density in units of the critical density, 
 which corresponds to a wavenumber of $k_B \simeq 0.1 (\Omega h) \impc$.
 Thus, for CDM-like models the range of the scales 
 we find for the break in the  APM,  $k=0.02-0.06 \impc$,
 implies $0.2 \simlt \Omega h \simlt 0.6$. 
 To be more precise we perform a $\chi^2$ fit to the CDM models in Bond \&
 Efstathiou (1984) using  the four APM $P(k)$ points 
 in the range $k=0.02-0.06 \impc$ to find $\Omega h \simeq 0.45 \pm 0.10$
 ($\Omega h = 0.2$ produces a $\chi^2 \simeq 9$, while $\Omega h = 0.4$
 gives $\chi^2 \simeq 1.3$). Thus the case $\Omega=1$ requires 
 $h \simeq 0.45 \pm 0.10$ while an open universe or one with a non-zero 
 cosmological constant, $\Lambda$, 
 can accommodate other values of the Hubble
 constant $h$. 
 For purely relativistic
 dark matter, like neutrinos,  the scale at which 
 the amplitude of fluctuations are damped is typically larger
 than for CDM, corresponding to the Hubble radius when the universe
 becomes non-relativistic. For these models the 
 measured break yields correspondingly larger values for $\Omega h$.

As shown in Figure \ref{apmsl},
the sharp change in the local slope of the APM  
between $k \simeq 0.05-0.1 \impc$ is not compatible with any 
CDM model, which have a broader peak.
Note that in Figure \ref{apmsl}, the results are 
independent of uncertainties in the overall normalization or in 
any linear bias that may be applied,
unlike Figure \ref{fig:pknonlin}.
We have also shown that non-linear evolution, 
is not sufficient to modify the shape of the 
linear CDM power spectrum to provide a good match to the 
shape of the observed APM spectrum.

We have argued in \S\ref{sec:varCDM} that simple variation of CDM models, such
us tilted or higher $\Omega_{B}$ models can not account for the APM 
observations.
Models in which a large fraction of the matter is relativistic
(such as Mixed Dark Matter) are more likely to match this type of 
sharp feature. 
The scale found here for the break, 
around $k \simeq 0.05 \impc$,
could give interesting constraints for these models.

\section*{Acknowledgements}
We thank George Efstathiou for supplying us with a 
copy of the $P^{3}M$ code.
We also thank Gavin Dalton, Radek Stompor and Carlos Frenk 
for help and stimulating discussions.
CMB acknowledges receipt of a PPARC research assistantship and 
support from CESCA (HCM/EC grant) during a visit to the
Institut d'Estudis Espacials de Catalunya.
EG acknowledges support from CSIC, DGICYT (Spain), project
PB93-0035 and CIRIT (Generalitat de
Catalunya), grant GR94-8001.

\bigskip

{\bf REFERENCES} 
\bigskip

\def\refe {\par \hangindent=.7cm \hangafter=1 \noindent}
\def\aj { ApJ, }
\def\apj { ApJ, }
\def\aa {A \& A, }
\def\ajs{ ApJS, }
\def\apjs{ ApJS, }
\def\mn { MNRAS, }
\def\apl { Ap. J. Let., }

\refe Bardeen, J. M., Bond, J. R., Kaiser, N., \& Szalay, A. S. 1986, \aj,
      304,15
\refe Baugh, C.M., Efstathiou, G., 1993, \mn 265, 145 (BE93)
\refe Baugh, C.M., Efstathiou, G., 1994, \mn 267, 323 (BE94)
\refe Baugh, C.M., Gazta\~{n}aga, E., Efstathiou, G., 1995, \mn 274,   
      1049 
\refe Baugh, C.M., Gazta\~{n}aga, E., 1996, \mn 280, L37 (BG96)
\refe Baugh, C.M., 1996, \mn 280, 267
\refe Bernardeau, F., 1995 A\&A 301, 309.
\refe Bond, J.R., Efstathiou, G., 1984, \aj, 285, L45
\refe Boschan, P., Szapudi, I., Szalay, A.S., 1994, \ajs 93, 65 
\refe Broadhurst, R.S., Ellis, R.S., Shanks, T.,
       1988, \mn 235, 827
\refe Bunn, E.F., Liddle, A.R., White, M., 1996, Phys. Rev. D., 54, R5917
\refe Cole, S., Fisher, K.B., Weinberg, D.H., 1995, \mn 275, 515
\refe Colless, M., Ellis, R.S., Taylor, K., Hook, R.N.,
      1990, \mn 244, 408
\refe Colless, M., Ellis, R.S., Broadhurst, T.J., Taylor, K., Peterson, B.A.,
      1993, \mn 261, 19
\refe Dalton, G.B., Efstathiou, G, Maddox, S.J., Sutherland, W.J., 1992, \aj
      390, L1
\refe Efstathiou, G., Bond, J.R., White, S.D.M., 1992, \mn 258, 1P
\refe Efstathiou, G., Davis, M., Frenk, C.S., White, S.D.M., 1985 \apjs 
      57, 241
\refe Efstathiou, G., Ellis, R.S.  Peterson, B.A.,
      1988, \mn 232, 431
\refe Efstathiou, G., Kaiser, N., Saunders, W., Lawrence, A., 
      Rowan-Robinson, M., Ellis, R.S., Frenk, C.S., 1990a, \mn 247, 10p
\refe Efstathiou, G., Sutherland, W.J., Maddox, S.J., 1990b, Nature, 348, 705
\refe Einasto, J., Einasto, M., Gottlober, S., Muller, V., Sarr, V., 
      Starobinsky, A.A., Tago, E., Tucker, D., Andernach, H., Frisch, P., 
      1997, Nature, 385, 139.
\refe Eke, V.R., Cole, S., Frenk, C.S., 1996, \mn 282, 263
\refe Ellis, R.S., Colless, M., Broadhurst, T., Heyl, J., Glazebrook, K., 
      1996,  \mn 280, 235
\refe Fall, S.M., Tremaine, S., 1977, \apj 216, 682
\refe Fisher, K.B., Davis, M., Strauss, M.A., Yahil, A., Huchra, J.,
      1993, \apj 402, 42
\refe Fosalba, P., Gazta\~naga, E., 1997 in preparation.
\refe Fry, J.N., Peebles, P.J.E. 1978, \aj 221, 19
\refe Fry, J.N., 1984, \aj 279, 499
\refe Fry, J.N., Gazta\~naga, E., 1993, \aj 413, 447
\refe Gazta\~naga, E. \& Baugh C.M. 1995, \mn 273, L1.
\refe Gazta\~naga, E. 1994, \mn 268, 913 (G94)
\refe Gazta\~naga, E. 1995, \aj 454, 561 (G95)
\refe Gazta\~naga, E., Bernardeau, F., 1997, \aa submitted , 
astro-ph/9707095.
\refe Gazta\~naga E., \& Frieman J.A., 1994, \aj 437, L13
\refe Gazta\~naga, E., Lacey, C. 1997, in preparation
\refe Groth, E.J., Peebles, P.J.E. 1977, \aj 217, 385
\refe Gunn, J.E., Weinberg, D.H., 1995, in Wide Field Spectroscopy and 
      the distant universe, proceedings of 35th Herstmonceux workshop.
\refe Goldberg, D.M., Strauss, M., 1997 \aj submitted astro-ph/9707209
\refe Hamilton, A.J.S., Kumar, P., Lu, E., Matthews, A., 1991, \aj 374, L1
\refe Jain, B., Mo, H.J., White, S.D.M., 1995, \mn 276, L25 
\refe Kaiser, N., 1984, \apj 284, L9 
\refe Kaiser, N., 1987, \mn 227, 1
\refe Kolb E.W., Turner M.S., 1990, {\it The Early Universe}, Addison-Wesley,
New York
\refe Limber, D.N., 1954, \apj 119, 655
\refe Loveday, L. , Peterson, B.A., Efstathiou, G.,
      Maddox, S.J., 1992 \aj 390, 338
\refe Loveday, L. ,  Efstathiou, G., 
       Maddox, S.J., Peterson, B.A., 1996 \apj 468, 1
\refe Maddox, S.J., Sutherland, W.J., Efstathiou, G.,   
      Loveday, L., Peterson, B.A. 1990a \mn 243, 692
\refe Maddox, S.J., Efstathiou, G., Sutherland, W.J.  
      1990b \mn 246, 433.
\refe Maddox, S.J., Efstathiou, G., Sutherland, W.J.  
      Loveday, L.,  1990c \mn 242, 43p
\refe Maddox, S.J., Efstathiou, G., Sutherland, W.J.  
      Loveday, L.,  1996 \mn 283, 1227
\refe Mo, H.J., Jing, Y.P., White S.D.M., 1997, \mn in press,
astro-ph/9603039.
\refe Peebles, P.J.E., 1980, {\it The Large Scale Structure of the 
      Universe:} Princeton University Press
\refe Press, W.H., Teukolsky, S.A., Vetterling, W.T., Flannery B.P., 1992,
      Numerical Recipes (Cambridge)
\refe Peacock, J.A., 1991, \mn 253, 1p
\refe Peacock, J.A., Dodds, S.J.,  1994, \mn 267, 1020
\refe Peacock, J.A., Dodds, S.J.,  1996, \mn 280, L19
\refe Rugers, M., Hogan, C.J., 1996, A.J., 111, 2135
\refe Saunders, W., Frenk, C., Rowan-Robinson, M., Efstathiou, G., 
      Lawrence, A., Kaiser, N., Ellis, R., Crawford, J., Xia, X.Y., 
      Parry, I., 1991, Nature 349, 32
\refe Seljek, U., Zaldarriaga, M., 1996, \aj, 469, 437
\refe Smith, C., Klypin, A., Gross, M., Primack, J., Holtzman, J., 1997, 
      MNRAS submitted, astro-ph/9702099 
\refe Stompor, R., Banday, A.J., Gorski, K.M., 1995, \mn 277, 1225
\refe Tadros, H., Efstathiou, G., 1996, \mn 282, 1381
\refe Tadros, H., \& Dalton, G.B. 1997, in preparation
\refe Tadros, H., 1996 Thesis, Univ. of Oxford
\refe Tytler, D., Fan, X.M., Burles, S., 1996, Nature, 381, 207
\refe Vogeley, M.S., Park, C., Geller, M.J., Huchra, J.P., 1992, \apj 391, L5
\refe White, S.D.M., Efstathiou, G., Frenk, C.S., 1993, \mn 262, 1023 
\refe Weinberg, D.H., 1994, Proceedings of the 35th Herstmonceux conference,
astro-ph/9409094. 

\end{document}